\documentclass[%
 reprint,
 amsmath,amssymb,
 aps,
]{revtex4-1}

\usepackage{graphicx}
\usepackage{dcolumn}
\usepackage{bm}

\usepackage{algpseudocode}  
\usepackage{amsthm}
\newtheoremstyle{algstyle}
  {5pt}
  {4pt}
  {\normalfont}
  {0pt}
  {\itshape}
  {.}
  {\newline}
  {}
\theoremstyle{algstyle}
\newtheorem{theorem}{Theorem}
\newtheorem{algorithm}[theorem]{Algorithm}

\usepackage[english]{babel}
\usepackage[utf8x]{inputenc}
\usepackage[T1]{fontenc}

\usepackage{mathtools}

\DeclarePairedDelimiter{\nint}\lfloor\rceil

\usepackage[colorlinks=false]{hyperref}

\begin{document}

\preprint{APS/123-QED}

\title{Efficient Design of Hardware-Enabled Reservoir Computing in FPGAs}

\thanks{\{bogdan.penkovskyi,laurent.larger,daniel.brunner\}@femto-st.fr}

\author{Bogdan Penkovsky}
\author{Laurent Larger}
\author{Daniel Brunner}

\affiliation{%
  FEMTO-ST / Optics Dept., UMR CNRS 6174, Univ. Bourgogne Franche-Comté,\\
  15B avenue des Montboucons, 25030 Besançon Cedex, France
}%

\date{\today}

\begin{abstract}
In this work, we propose a new approach towards the efficient optimization and implementation of reservoir computing hardware
reducing the required domain expert knowledge
and optimization effort.
First, we adapt the reservoir input mask
to the structure of the data
via linear autoencoders.
We therefore incorporate the advantages
of dimensionality reduction
and dimensionality expansion achieved by
conventional algorithmically efficient
linear algebra procedures
of principal component analysis.
Second, we employ evolutionary-inspired
genetic algorithm techniques
resulting in a highly efficient
optimization of reservoir dynamics with dramatically reduced  number
of evaluations comparing to exhaustive search.
We illustrate the method on
the so-called single-node
reservoir computing
architecture, especially suitable
for implementation in ultrahigh-speed hardware.
The combination of both methods
and the resulting reduction of
time required for performance optimization of a hardware system
establish a strategy
towards machine learning hardware
capable of self-adaption to optimally solve specific problems.
We confirm the validity of those principles
building reservoir computing hardware based on a field-programmable
gate array.

\end{abstract}

\pacs{Valid PACS appear here}
\maketitle


\section{Introduction}

Machine learning (ML) development has drastically progressed during
the last decade. To only name a few examples,
now machines can accurately
describe images \cite{LeCun2015},
identify and recognize faces \cite{amos2016openface},
recognize speech \cite{Graves2013a}
and compose music \cite{Johnson2017,Johnson2017b}.
In 2017,
AlphaGo Zero and AlphaZero with no prior
domain knowledge have beaten the best human
and machine players both
in Go \cite{Silver2017} and chess \cite{Silver2017a} games.
These all are challenges which, until recently,
were thought to remain reserved for the human intelligence only.

However, the efficiency of current ML methods is restricted
by hardware, which in its turn
is fundamentally limited by the minimal transistor size.
Another potential issue
is related to the fact the vast majority
of ML hardware 
rely only on a single design:
the Turing-von Neumann architecture.
That results in the second, conceptual limitation:
our machines
are constrained by their implementation's design, and that design is mostly one.
A viable way to circumvent present limitations is by
shifting the design paradigm away from Turing–von Neumann architectures.
This shift may also give  insight into questions
related to self-adapting hardware.

In this paper, we demonstrate the implementation of a single-node reservoir computer (RC) \cite{Appeltant2011} in
field-programmable gate array (FPGA) hardware. Hardware-implementation of such time-delay reservoirs (TDRs) is particularly resource efficient. They mostly consist of a first-in first-out (FIFO) memory combined with a single nonlinear dynamical node. Yet even such simple systems have a parameter space with dimensionality too-high for exhaustive parameter optimization to be realistic. We propose genetic algorithms (GAs) for enabling our TDR-hardware to
quickly optimize its dynamical properties, achieving a reduction by 2-3 orders
of magnitude in optimization effort. Thanks to such improvement and the ease of hardware-implementation of a GA, future systems will be able to adjust
to unforeseen changes in data~\cite{Torresen2004}.
Addressing potential bottlenecks in the input-interface, we merge the system's input weight-matrix with auto-encoders realizing principle component analysis. With this approach we achieve data-injection efficiency increase by a factor of 1.8.
As a proof-of-concept, we prototype the self-adapting
system in simulation, evaluate it on a speech recognition benchmark,
and verify the validity of the approach using an FPGA.
The results achieved by FPGA-based RC with limited bit resolution
closely match those obtained in simulations with
noise, confirming our strategy for future design of TDR-systems based on physical nonlinear substrates.

\subsection{Reservoir computing. Single-node approach}

\begin{figure}[ht]
\centering
\includegraphics[width=\columnwidth]{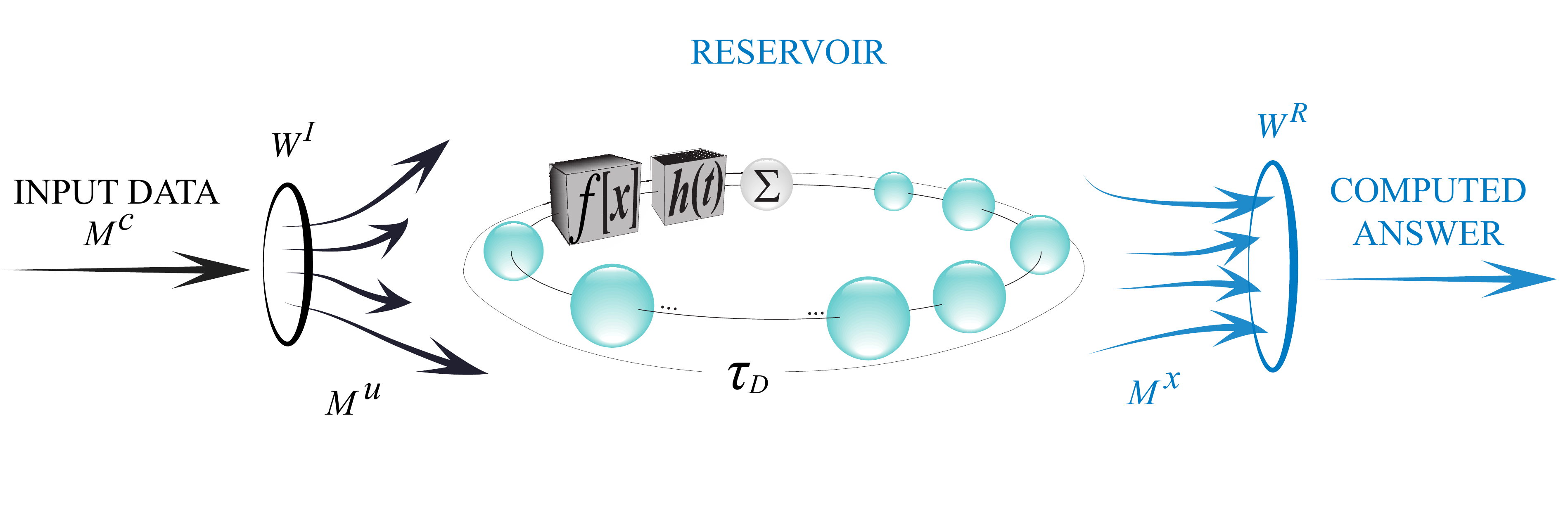}
\caption{\label{fig:rc-single-node}
\textbf{Single-node reservoir computing architecture}.
The three core components are:
input masking,
reservoir transformation,
and linear readout.
First, the input data $M^c$ are
masked by multiplying
the mask $W^I$,
then the masked input $M^u$ is transformed by
reservoir's nonlinear delay dynamics.
Finally, the answer is obtained
by multiplying the readout $W^R$
and the reservoir's state $M^x$.
}
\end{figure}

Reservoir computing (RC) first appeared as a modification
to recurrent neural networks (RNNs) training
and was proposed independently in
\cite{Jaeger2001, steil2004, Maassa}.
Due to the retained internal state, RNNs are also known
as ``deep'' neural networks.
Because of the practically infinite number of hidden layers,
gradient-based training of RNNs often suffers from
exploding or vanishing gradients.
RC solved the problem of RNN training by applying
the principle of a random mapping.
Reservoir acts as a spatiotemporal kernel
mapping the low-dimensional
input information onto a higher-dimensional state space,
where this information is expected to become linearly
separable~\cite{Hermans2011}.
Therefore, a linear readout interpreting \textit{transient}
dynamics in the high-dimensional network's state
should be sufficient to interpret that information.
As a consequence, instead of adapting
the whole recurrent network, only the linear readout
layer is trained in RC.

The RC approach achieves multiple objectives:
(1) the training procedure is fast, and
(2) is guaranteed to converge
using conventional linear
algebra techniques, and crucially for the development of novel computing systems,
(3) the fixed nonlinear part of the network can practically be delegated
to low-level hardware,
i.e. physically existing dynamical systems,
not limited to digital electronics.
Reservoir computing is a
computation paradigm that potentially
addresses the issues of
inherently fast and energy-efficient hardware.
This is mainly due to its
support of information processing directly on the 
very hardware level. 
Several experimental implementations of RC hardware are known: using
digital-analog electronics \cite{Appeltant2011, Soriano2015},
electro-optical and all-optical systems 
\cite{Larger2012,Duport2012,Brunner2013,Vandoorne2014}, and
spintronic nanoscillators \cite{Torrejon2017}.
Both numerical and experimental RC systems often
beat state-of-the-art in speed
(such as in speech recognition task \cite{Larger2017})
and accuracy
(e.g. time series prediction \cite{Jaeger2004}).


The single-node node RC is 
a technique that uses a complex nonlinear delayed-feedback system as a
dynamical reservoir~\cite{Soriano2013}.
This delay system is a recurrent network which retains its
internal state as the state of the delay line.
The single-node 
approach to RC takes advantage of delay dynamics,
which has a high-dimensional, mathematically speaking even
infinite-dimensional phase space, and can be interpreted
as a virtual network \cite{Appeltant2011,Larger2013a}.
Single-node RC
is frequently implemented in hardware as it is a 
technologically efficient way to construct a nonlinear reservoir network.
The benefit of the method is the
ability to reduce the physical neural
network's size to a single nonlinear unit,
thereby resulting in a smaller number of dynamical
parameters to control.
Moreover, the single-node RC architecture is especially
suitable for ultrahigh-speed photonic hardware
implementations~\cite{Larger2017}.
Numerous other demonstrators have
validated the single-node RC approach
showing autonomous RCs \cite{Antonik2017a} and 
variety of hardware architectural and training modifications
\cite{Martinenghi2012,Hermans2015b,Argyris2018PhotonicCommunications}.

The present work is two-fold. First, it
addresses the problem of 
\textit{hyperparameters}~\footnote{By hyperparameters we mean those parameters which are set before the learning begins.}
optimization in RC,
which was also discussed in \cite{Yperman2016BayesianComputing}.
However, we apply genetic algorithms as the optimization
strategy and validate the method on actual hardware,
addressing the possibility of real-world hardware design. In \cite{Ferreira2009GeneticOptimization,Rigamonti2016EchoEngine}
genetic algorithms were also applied, however to a different
RC architecture (echo state network) and with no attempt at hardware implementation, while
in this work we consider the single-node architecture.
Second, an optimal input matrix is constructed
by stripping input data of trivial information.
The combination of both methods, genetic algorithm and  \textit{autoencoder},
is applied for the first time to hardware-implemented RC.

A general single-node RC architecture
is schematically represented in Fig. \ref{fig:rc-single-node}.
Input information $M^c$ is masked by
the input mask $W^I$ and then, mapped
as $M^u$
on a high-dimensional state space
of a delay reservoir.
Then, the reservoir's nonlinear
response creates a state matrix
$M^x$. The final answer is obtained by a linear readout,
i.e. by multiplying matrix $W^R$.
An introduction to the single-node 
RC can be found in \cite{Appeltant2011, Schumacher2015, Larger2017}.

\subsection{Input streamlining}

In \cite{Miikkulainen1997} it was suggested
that information processing in the brain
(e.g. in the primary visual cortex) is performed
in three stages:
first, input projection into
principal feature dimensions,
second, redundant information filtering,
and finally third, higher-level information
processing.
Motivated by that strategy,
we propose automatic feature weighting
via redundant information filtering,
to enhance the conventional
random masking of RC. That is achieved
by principal component analysis (PCA),
a technique that constructs linear combinations
of input features.

First, we apply PCA to remove
the dimensions with the lowest variance,
i.e. to compress
the input data. That allows us to focus on the
input data's most relevant structure.
Then, an inverse to compression
operation, dimensionality expansion, is 
employed to restore the shape of the inputs.
These two linear operators,
compression and expansion~\footnote{Linear operators
are expressed as matrices. The relationship between compression
and expansion (decompression) obtained after PCA is a matrix transpose operation.},
partially remove irrelevant feature information,
such as noise.
Therefore, PCA plays the role of an
\textit{autoencoder} (autoassociative neural network),
a network utilized to learn efficient data coding.
In practice, application of PCA-based compression-decompression
is very similar to the application of an autoassociative neural network with a single hidden layer \cite{Bishop2006}.
Finally, random 
masking conventional to RC
is performed in order
to map the information
onto a higher dimensional
state space of the reservoir.

\subsection{Self-adapting reservoir dynamics}

Due to the simplified training step in RC,
the main action in system performance optimization is dynamical system parameters exploration.
While the single-node RC method's complexity is much reduced,
this can still be a substantial bottleneck
for real-world applications
since each problem may require a different set of dynamical
parameters. For instance, a set of
optimized speech recognition
system parameters
is potentially different from that of handwriting recognition.
Therefore, quick parameter search is crucial when adapting
RC to a new task.

Another case when dynamical parameters optimization
is essential is testing new RC substrates when there is no
prior RC parameter estimate. This becomes even more relevant
when choosing between several alternative dynamical RC systems
that differ in materials. For example, performing speech recognition
in a bucket as in \cite{Fernando2003},
which liquid is more suitable?
A mixture, if then in which proportions, at which temperature,
how deep would be an optimal reservoir?

As illustrated, one typically deals with multi-dimensional
hyperparameter optimization.
There are two quite contrary approaches towards 
RC hyperparameter optimization which are currently prevailing.
One is a so-called hyperparameter fine-tuning
(essentially, trial and error)
method. Here, one relies on often partially heuristic arguments why a certain set of starting hyperparameters might be well suited. From that point one searches for the nearest, potentially only local, performance optimum.
Although this ad-hoc
practice can be frequently observed in the ML community,
it does not guarantee an optimal hyperparameter combination.

An opposite case is
a more systematic optimization approach,
the grid search (GS) technique. The method consists
in an exhaustive search of all hyperparameter combinations
under certain constraints.
GS  has its advantage in guaranteeing
the identification of a global
hyperparameter optimum, provided that the parameter search grid
is sufficiently dense.
In addition, the technique provides a multidimensional error landscape,
giving insight into the structure of the parameter
space and through that potentially into the relationship
between task and computing system.
However, GS may take substantial optimization time
due to the exponentially increasing amount of data points with each
additional optimized hyperparameter. Therefore, in practice GS is often
limited to three-four scanned hyperparameters.
As a result, a high number of optimized hyperparameters in modern ML
applications renders GS utilization inefficient or even virtually impossible.

In the present work, we
provide a strategy how to create
a data-driven self-adapting
reservoir dynamics by employing an evolutionary selection-inspired
technique known as genetic algorithm (GA) \cite{Deb1999}.
The GA optimization method can
be regarded as a sweet spot between the
two optimization extrema:
it provides a systematic search while dramatically
reducing the number of trials, making
GA especially suitable
for RC hardware design.
The algorithm works with a population of \textit{chromosomes}
encoding RC dynamics properties and thereby,
works with a population of RC models.
The evolutionary process is achieved by applying so-called
\textit{genetic operators}
recombining the information in chromosomes.
Every population is evaluated
so that the most fit chromosomes, i.e. describing the
most useful reservoir dynamics, tend to survive and reproduce,
leading therefore to a better average configuration in every new population. Comparing to classical
optimization methods such as gradient descent,
(1) GAs tend to operate on encodings,
not the actual parameter values,
(2) GAs can handle problems
with both continuous and discrete search spaces,
and (3) at no additional computational cost GAs produce several alternative configurations to choose from.

\section{Methods}

\subsection{Reservoir computing}

Prior to the reservoir transformation (Fig. \ref{fig:rc-single-node}),
each input data matrix $\mathbf{M}^c$,
consisting of $L$ input feature vectors
$\mathbf{c}(n), n = 1 \dots L$,
is first
masked by multiplying an input mask
$\mathbf{W}^I_2$ and then, temporally encoded:

\begin{equation}
\begin{array}{lll}
\mathbf{W}^{I}_2\mathbf{c}(n) & = & \mathbf{W}^{I}_2\left(c_{1}(n),c_{2}(n),\dots,c_{M}(n)\right)\\
 & = & \left(u_{1}(n),u_{2}(n),\dots,u_{N}(n)\right)\\
 & = & \left(u(t+\theta),u(t+2\theta),\dots,u(t+N\theta)\right),
\end{array}\label{eq:rc-masking}
\end{equation}

\noindent where
$M$ is the input data dimensionality, and
$N$ is the reservoir network size.
$\mathbf{W}^{I}_2$ is calculated
from Eq.~\eqref{eq:wI2}
with
$\mathbf{W}^{I}\in\mathbb{R}^{N\times M}$, $(N > M)$ having
weights randomly drawn
from $\left\{ -0.4; 0; 0.4 \right\}$ 
(30\% connectivity) and
remaining fixed for all experiments.
Finally, the temporally-encoded input signal $u(t)$
is kept constant in-between 
times $(t + i \theta, t + (i+1) \theta)$,
$i = 1 \dots N$,
corresponding to the temporal separation between
the virtual nodes \cite{Appeltant2011}.
This temporal encoding technique is
sometimes called a \textit{sample-and-hold} operation.

The temporal information input signal $u(t)$
is subsequently processed by the
delayed-feedback nonlinear
system reservoir (Eq. \eqref{eq:rc-lowpass}).
The choice of this particular
reservoir dynamics model
was motivated by its recent
implementation as a 
substrate for numerous photonic RC
devices~\cite{Appeltant2011, Brunner2013, Larger2017}, and can often be described by the low-pass delay-differential equation

\begin{equation}
\tau\dot{x}(t) = -x(t) + f\left(x(t-\tau_D)+\rho u(t)\right).
\label{eq:rc-lowpass}
\end{equation}


\noindent In our case, we employ $f(x) = \beta \sin^2(x+\Phi_0)$ as nonlinearity.
The nonlinear dynamics parameters
$\tau$, $\beta$, $\Phi_0$, and $\rho$ are subject to
optimization while delay time $\tau_D=6$ is kept constant in our experiments. Moreover, similar bandpass-filtered
systems can be easily implemented
in electro-optical substrates~\cite{Larger2015b}.

The result of RC $\mathbf{y}(n)$ is
computed as:
\begin{equation}
\mathbf{y}(n) = \mathbf{W}^R \mathbf{x}(n),\label{eq:rc-readout}
\end{equation}
where vector
$\mathbf{x}(n)=\left(x(t+\theta),x(t+2\theta),\dots,x(t+N\theta)\right)$
is the decoded nonlinear reservoir response
(Eq.~\ref{eq:rc-lowpass}).
The linear readout weights $\mathbf{W}^R$ are obtained
on a computer from previously processed
data samples (Eqs~\eqref{eq:rc-masking}-\eqref{eq:rc-lowpass})
using the ridge regression:

\begin{equation}
\mathbf{W}^R=(\mathbf{M}_x\cdot \mathbf{M}_x^{\intercal}+\lambda \cdot \mathbf{I})^{−1}(\mathbf{M}_x\cdot \mathbf{T}^{\intercal}),\label{eq:ridge-regression}
\end{equation}

\noindent where $\lambda \ll 1$ is a
small regularization constant.
$\mathbf{M}_x \in \mathbb{R}^{N \times Q}$ is a feature matrix of concatenated horizontally state vectors
$\mathbf{x}(n)$.
$\mathbf{T}  \in \mathbb{R}^{K \times Q}$ is a teacher matrix,
$Q$ denotes output dimensionality,
and $K$ the number of training feature vectors.
For classification tasks, the teacher is a
\textit{one-hot encoded} matrix,
i.e. consists of
target answer vectors 
$\mathbf{y}^{\rm tgt} \in \mathbb{R}^{K \times 1}$ 
where the only nonzero elements
correspond to the correct class label.

\subsection{Enhanced masking via PCA}

In our approach, we adapt the input mask
to the data structure particular to each task.
A new input matrix $\mathbf{W}^I_2$
is constructed as a superposition
of three linear operators,
compression $\mathbf{W}_c$,
decompression $\mathbf{W}_c^{\intercal}$,
and conventional
(random) masking $\mathbf{W}^I$:

\begin{equation}
\mathbf{W}^I_2 = \mathbf{W}^I \cdot \mathbf{W}_c^{\intercal} \cdot \mathbf{W}_c,
\label{eq:wI2}
\end{equation}

\noindent where the transposed matrix pair 
$\mathbf{W}_c$ and $\mathbf{W}_c^{\intercal}$
is calculated using the standard 
unsupervised dimensionality reduction
technique of
principal component 
analysis~\cite{Pearson1901, Hotelling1933};
$\mathbf{W}^I$ is randomly generated
as usual for RC.
The resulting mask $\mathbf{W}^I_2$
remains fixed during all RC experiments
for the given task. Furthermore, as it optimizes input information content, it is not optimized for a particular set of dynamical reservoir dynamics.

Principal component analysis (PCA) can be described as follows.
First, an isolated subset of data is selected
such that it reflects the data distribution of the whole dataset.
Then, a covariance matrix $\mathbf{\Sigma} \in \mathbb{R}^{M \times P_0}$
is constructed using said subset. Here,
$M$ is the input dimensionality and $P_0$ is
the total number of 
feature vectors in the subset.
During the next step, a new matrix $\mathbf{W} \in \mathbb{R}^{M \times M}$
is obtained such that columns in $\mathbf{W}$ are eigenvectors
of $\mathbf{\Sigma}$ sorted by
decreasing magnitude of corresponding eigenvalues.
This is achieved via \textit{singular value decomposition}.
Finally, a compression matrix $\mathbf{W}_{c}\in\mathbb{R}^{M' \times M}$
is constructed 
by selecting the first $M' < M$
vector-columns of $\mathbf{W}$,
the principal components,
and transposing the resulting matrix.

The superposition of compression $\mathbf{W}_c$
and decompression $\mathbf{W}_c^{\intercal}$
operators is an autoencoder.
The autoencoder
$\mathbf{W}_c^{\intercal} \cdot \mathbf{W}_c$
facilitates general
data structure extraction by learning the most relevant features,
while $\mathbf{W}^I$
helps to map the input data onto a higher
dimensional state space.
Therefore, the new operator $\mathbf{W}^I_2$
can be interpreted as a mask made more sensitive to the most
relevant features in the input data,
rather than the commonly employed simple random feature mapping via $\mathbf{W}^I$.


Another implication for  ML hardware implementation
of our architecture 
is tackling the input bottleneck.
By decomposing $\mathbf{W}^I_2$
into two independent steps, first
compression  $\mathbf{W}_c$
and second masking and decompression
$\mathbf{W}^I \cdot \mathbf{W}_c^{\intercal}$,
both steps can be performed
by different units.
One can therefore preprocess the information by a simplistic special unit according to the first step. The information ultimately to be injected into the physical reservoir is then
compressed at ratio $M / M'$.

%

\subsection{Hyperparameter self-optimization}

Introduced in \cite{Holland1975AdaptationIntelligence},
genetic algorithms (GAs) are a family
of evolutionary-inspired techniques
based on the idea of survival of the fittest.
Genetic algorithms can be generally described as follows:

\begin{algorithm}
\begin{algorithmic}
\State {}
\State Random initial population
\Repeat
\State Fitness evaluation
\State Selection
\State Crossover and mutation
\Until{Converged}
\end{algorithmic}
\end{algorithm}

First, an initial random population of chromosomes
is generated. This stage corresponds to a completely
random search.
Then, each chromosome is evaluated
according to a certain loss function.
The objective of the optimization is to minimize
the loss function, therefore the most
fit are individuals with the lowest error score.
The most fit individuals are more likely
to be selected for reproduction.
Finally, application of genetic operators
of \textit{crossover} and \textit{mutation}
over selected chromosomes results in a new population of chromosomes with improved average fitness.
The process is repeated until convergence
or for a fixed number of iterations.
Note that GAs use probabilistic computations
and each realization may lead to a different
result.
Below, we provide more details
specific to our particular
GA implementation~\footnote{The employed genetic algorithms library was https://hackage.haskell.org/package/GA}.

\subsubsection*{Hyperparameters encoding}

A chromosome represents a unique hyperparameters configuration,
and therefore encodes all optimized hyperparameter values.
Here, we restrain ourselves to optimize only RC dynamics parameters,
however, any other hyperparameters,
such as for example the number of principal components,
could be potentially encoded in a chromosome.
To describe the encoded parameter values, we utilize
a binary encoding. The advantage of this scheme
is a straightforward implementation of genetic
operators. The shortcoming is an always finite resolution
of parameters encoded in chromosomes.
Note that this shortcoming 
is not relevant to physical RC realizations
since physical systems usually do not allow very
precise dynamical parameters tuning.

For the sake of illustration let us decode a
10 bit binary chromosome \texttt{<11.000|00.110>}.
In this example, the values are represented
as signed 5 bit fixed-point numbers with a 3 bit fractional part.
For convenience,
we have separated the values with a '\texttt{|}'
and fractional parts with a '\texttt{.}'. Hence we see that there
are two values encoded with the resolution of $2^{-3}$.
Minimal and maximal possible encoded values are $-2$ and $1.875$.
By convention, the first (sign) bit signifies
the maximal power of two with a negative value, here it
is $-2^1$.
The subsequent bits have positive values and correspond to
powers $2^0$, $2^{-1}$, $2^{-2}$, and $2^{-3}$, respectively.
Therefore, the first encoded value $\texttt{<11.000>} = -2^1 \cdot 1 + 2^0 \cdot 1 + 2^{-1} \cdot 0 + 2^{-2} \cdot 0 + 2^{-3} \cdot 0 = -1$.
The second value 
$\texttt{<00.110>} = -2^1 \cdot 0 + 2^0 \cdot 0 + 2^{-1} \cdot 1 + 2^{-2} \cdot 1 + 2^{-3} \cdot 0 = 0.75$.
Note that we vary parameter ranges and bit resolutions
between actual parameter values (see Table \ref{tab:par_ranges}).

\subsubsection*{Fitness evaluation and reproduction}

Before genetic operators can be applied,
the chromosomes in population have to be evaluated against
a certain loss function.
Therefore, an $N_{pop}$ (see Table~\ref{tab:ga_par})
number of RC systems are
trained with dynamics parameters encoded in corresponding chromosomes.
RC systems are then evaluated on an isolated validation dataset.
This allows us to assign losses in terms of word error rates.

\begin{table}
\centering
\begin{tabular}{l|r}
GA meta-parameter & Value \\\hline
Crossover rate, $r_c$ & 88\% \\
Mutation rate, $r_m$ & 12\% \\
Crossover prob. (uniform crossover) & 50\% \\
Bit mutation prob., $p_u$ & 20\% \\
Population size, $N_{pop}$ & 20 \\
Archive size, $N_{ar}$ & 12 \\
Total no. of generations & 40
\end{tabular}
\caption{\label{tab:ga_par} Summary of GA parameters. 
Total no. of RC training: $40 \times 20 = 800$ evaluations.}
\end{table}

A tournament selection is applied when selecting parent
chromosomes for reproduction.
The selection is performed using genetic material
from current population
with $N_{pop}$ members and a certain number of the fittest chromosomes
(``archive'') $N_{ar}$ from the previous population.
Therefore, the total number of chromosomes for
tournament selection is $N_{pop} + N_{ar}$.
The selection mechanism first randomly choses two chromosomes.
Then, only the chromosome with the smaller loss (a tournament ``winner'') is
selected as a parent.
The crossover operator is binary and therefore requires two parents, i.e. two tournaments are performed before a crossover.
Whereas only a single parent is needed for mutation.
Naturally, the same chromosome may take part in several tournaments.

The crossover operator recombines information contained
in all selected parent chromosome pairs
and $\nint{r_c \cdot N_{pop}}$ new chromosomes are created.
Here, $\nint{\cdot}$ denotes ``nearest integer'' (rounding).
A child chromosome is generated by taking each bit with the
probability of 50\% from the first parent or from the second parent otherwise.
In the literature this is called a \textit{uniform crossover} operator which is inspired by the chromosomal crossover in biological cells.

Mutation creates $\nint{r_m \cdot N_{pop}}$ new chromosomes.
During mutation, each chromosome bit is flipped with the probability $p_u$.
It is crucial to balance
mutation rate $r_m$ and mutation probability $p_u$. 
If they are too high,
the obtained information so far may not have been used properly.
If too small, the algorithm may converge prematurely.

\subsection{Hardware implementation}

FPGA was chosen as an experimental platform
for RC hardware development due to several reasons.
First of all, FPGAs are well controllable
and re-programmable. Second, by leveraging reservoir computing,
FPGAs allow for a real-time prediction at MHz rate \cite{Penkovsky2017,Canaday2018RapidComputer}.
Third, existing
FPGA design may be ported to even faster electronic
hardware such as application specific integrated circuits
(ASICs).
We decided to implement asynchronous
communication between FPGA and external
world, as well as between FPGA modules.
A three wire communication protocol
was implemented~\cite{Penkovsky2017}.
This should facilitate applications
where data are coming asynchronously,
such as communication with remote devices in
a larger network.
Moreover, having a common asynchronous communication protocol
improves composition modularity and, as a result,
facilitates isolated module verification.
We investigate time-delay reservoirs since
combining both space and time multiplexing,
i.e. having multiple physical nodes with delays,
has potential implementing extremely high-dimensional
reservoirs~\cite{Canaday2018RapidComputer}.

\begin{figure}[h]
\centering
\includegraphics[width=\columnwidth]{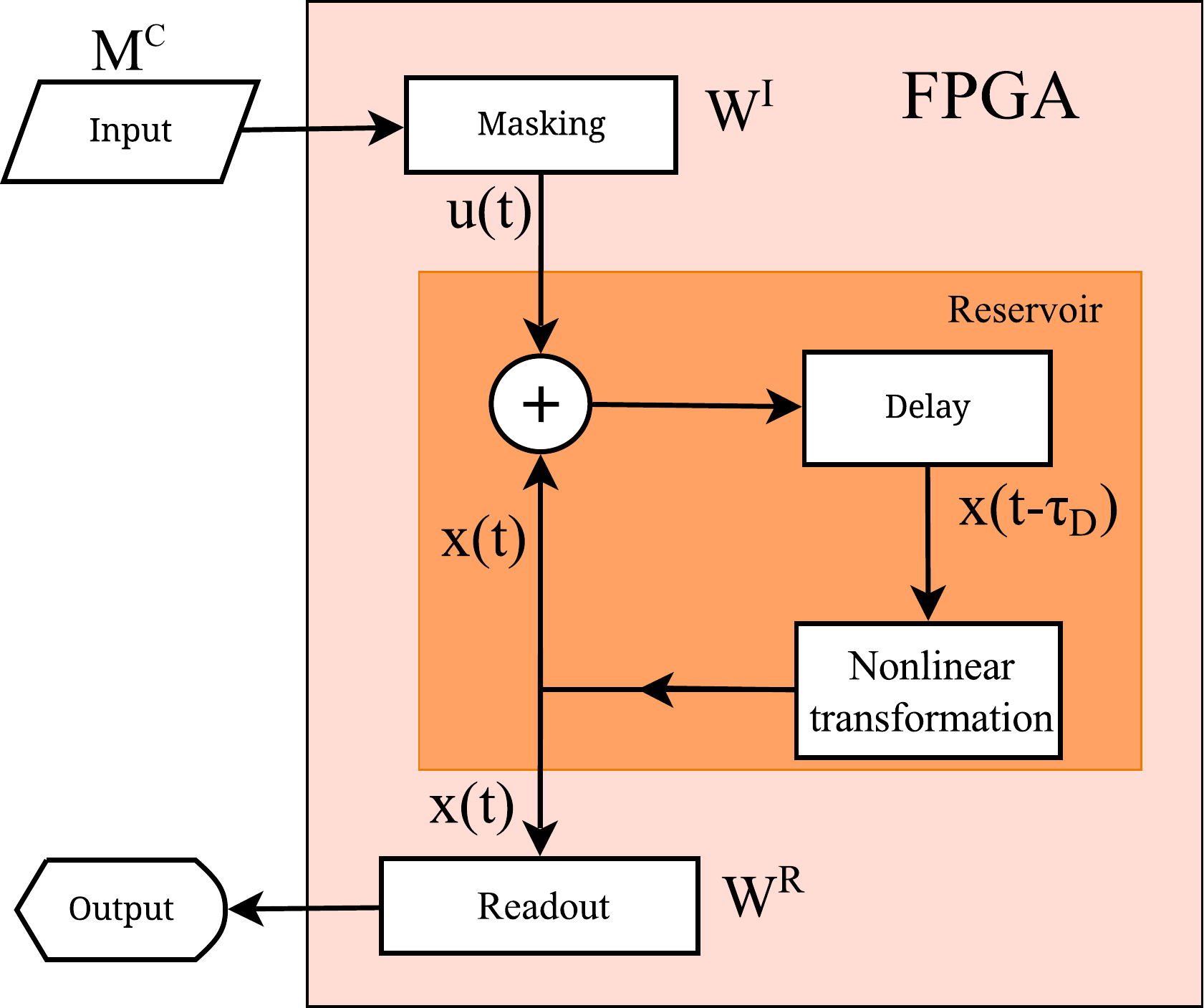}
\caption{\label{fig:fpga-arch}
\textbf{FPGA-based standalone RC architecture}
implements all essential RC blocks:
masking, reservoir,
and readout (cf. Fig. \ref{fig:rc-single-node}).
Arrows represent
16+2 bit wide asynchronous communication buses.
}
\end{figure}

We employ an 
Artix-7 (XC7A100T)
FPGA chip as a digital hardware
substrate for RC.
The major part of hardware design is
done in a hardware-synthesizable subset of
Haskell language known as Clash project~\footnote{Clash compiler is available via http://clash-lang.org}.
This hardware design approach
achieves two goals:
First, both hardware (FPGA) and software (RC simulation)
have a shared environment.
Certain pieces of code such as the Heun's integration
scheme implementing Eq.~\eqref{eq:rc-lowpass}
are simply reused by the FPGA design.
Second, the high-level functional language
drastically simplifies the hardware design and
verification workflow.
Clash compiles the design into a low-level VHDL hardware
description language.
Finally, Vivado Design Suite
generates the bitstream which directly
configures the FPGA.

The implemented architecture (Fig. \ref{fig:fpga-arch})
is a pipeline of three components working in parallel:
masking, delayed-feedback dynamics, and readout.
First, the information
input is compressed
on a computer using matrix $\mathbf{W}_c$ precalculated
via PCA. The resulting
data are
transferred to the FPGA
via a USB cable
using the serial UART protocol.
The information input block 
is implemented on FPGA
and fuses both
masking and decompression
$\mathbf{W}^I \cdot \mathbf{W}_c^{\intercal}$
as a single
matrix-vector multiplication operation.
The matrix-vector multiplication is real-time,
i.e. the component is instantly available after
previous vector has been input,
and is based on MAC (multiply-and-accumulate) circuits
conforming to the asynchronous communication protocol.
The data are then transferred
to the reservoir block
which simulates the delay dynamics of
Eq.~\eqref{eq:rc-lowpass}
using the second-order Heun's method.

The FPGA implements the 16-bit fixed-point
arithmetic, thus introducing
quantization (digitization) noise.
The impact of quantization noise
could be strongly reduced by an implementation based on
a floating-point module.
However, the downsides of the floating-point
FPGA implementation are more consumed 
programmable logic area
and potentially slower processing rates.
To demonstrate the practical applicability
to other RC realizations,
we stick to the less accurate fixed-point
representation natively
supported by our hardware.

During the training step, the system is run without the 
readout component.
The resulting dynamics is sent
to a computer where the readout matrix $\mathbf{W}^R$
is obtained with Eq. \eqref{eq:ridge-regression}.
During the testing step, to avoid 
model overfitting, we utilize a separate
testing dataset,
i.e. data neither used in FPGA training,
nor in model optimization.

\section{Results}

\begin{figure*}[ht!]
\centering
\includegraphics[width=17cm]{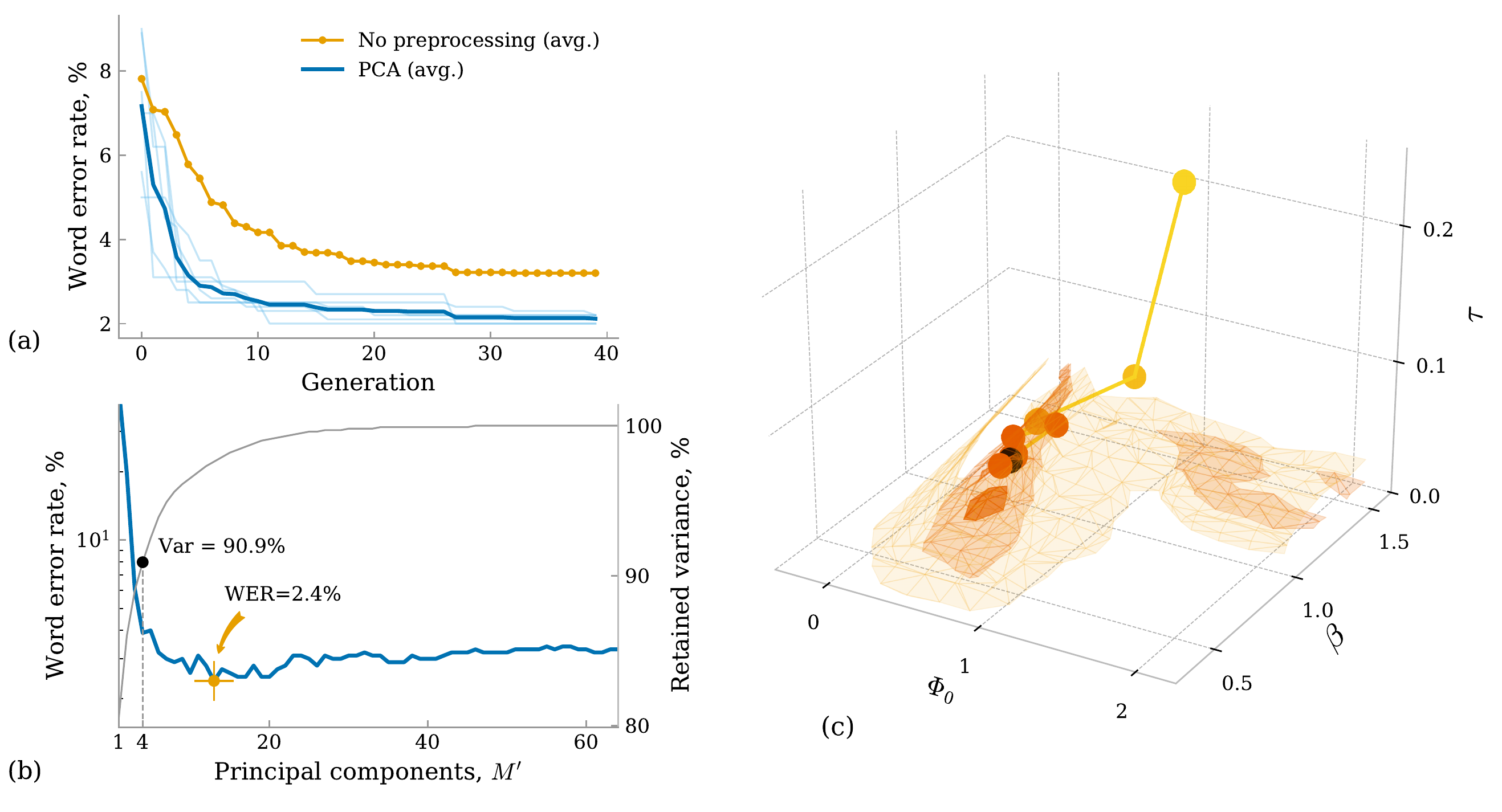}
\caption{\label{fig:pca_ga}
(a) 
\textbf{Genetic algorithm convergence averaged
over 6 individual runs}.
Individual genetic algorithm (GA) runs with
PCA preprocessing are illustrated by narrow
light-blue lines.
The PCA preprocessing not only provides better accuracy
(bold blue line)
but also faster GA convergence 
and better overall performance than
without PCA (orange dotted).
Both experiments were conducted for a network of $N=600$ virtual neurons.
(b)
\textbf{Selecting the number of principal components}.
Principal component analysis predicts four
as the minimal principal components 
number for current dataset. That corresponds
to the retained variance of 90.9\%.
The most accurate
value $\text{WER}=2.4\%$ is achieved at 13 principal components (orange crosshair)
with 97.6\% retained variance.
The small number of principal components
($M'=13$) compared to
the original number of channels ($M=64$)
indicates that the space containing human speech is sparse.
Fixed (suboptimal) dynamical parameters are
$\tau=5 \cdot 10^{-3}$, $\tau_D=6$, $\beta=0.8$, $\Phi_0=0.3$, $\rho=1.5$
(Eq. \eqref{eq:rc-lowpass}).
(c) \textbf{Projection of multidimensional
error surfaces in 3D parameter space}.
The fixed parameter is $\rho=1.5$.
The error landscape in 
the three parameter dimensions
($\Phi_0$, $\beta$, $\tau$)
is characterized by extensive grid search
(11,340 data points).
An example evolution of
the best chromosomes in each GA
generation is visualized with circles
converging to a local error 
minimum after $\sim 20$ generations.
The final parameter obtained by this GA run is 
$\text{WER}=2.5\%$ marked in black color.
The nested error isosurfaces
correspond to $\text{WER}=10,5,3\%$,
respectively. The darkest orange volume
contains the absolute error minimum
of $\text{WER}=1.9\%$.
}
\end{figure*}

\begin{table}
\centering
\begin{tabular}{c|ll|c}
Parameter & Min value & Max value & Resolution step \\\hline
$\tau$ & $7.8 \cdot 10^{-3}$ & $0.99$ & $2^{-7}$ \\
$\beta $  & $-4$ & $3.98$ & $2^{-6}$ \\
$\Phi_0$  & $0$ & $\pi$ & $2^{-6}$ \\
$\rho$  & $-4$ & $3.88$ & $2^{-3}$
\end{tabular}
\caption{\label{tab:par_ranges} Parameter ranges used
for GA search}
\end{table}

To benefit from the underlying
recurrent network,
we apply RC to a time-dependent
signal, human speech.
The benchmark employed in this paper is a speech recognition
task based on the clean isolated digits subset
of Aurora-II database \cite{Hirsch2000} (2412 samples).
Following the established speech recognition paradigm,
we model the dynamics of the inner ear and utilize 
Lyon model \textit{cochleagrams} \cite{Slaney1988}
as 64-dimensional inputs to the reservoir described
by Eqs \eqref{eq:rc-masking}-\eqref{eq:rc-lowpass}.

We start our experimentation with unoptimized
reservoir dynamics parameters and perform GA search.
Table \ref{tab:par_ranges} summarizes parameter ranges
selected with respect to \textit{physically meaningful}
RC dynamics. For instance, the delay system
Eq. \eqref{eq:rc-lowpass} is $\pi$-periodic because of
the nonlinear function $f(x)=\sin^2(x)$, therefore
we restrict $\Phi_0 \in [0; \pi]$;
parameter $\tau$ cannot be large with respect to delay time $\tau_D$,
otherwise the system's complexity is substantially reduced;
finally, $\beta$ and $\rho$ cannot be large otherwise the system will
bifurcate away from useful dynamics.
Otherwise, we do not provide
any knowledge common to RC implementations (such as edge-of chaos),
i.e. dynamics parameters are
self-adapted to the
speech recognition task.


To evaluate the classification accuracy
during GA and GS optimizations, a so-called
two-fold cross-validation is employed.
First, training is performed on a group of 500 digits
and another group of 500 digits is used for validation.
Then, the roles are reversed, i.e. training is performed on
the second group and validation, on the first one.
Finally, a separate dataset of 1000 digits is used
for testing the FPGA implementation.
The remaining 412 samples are employed for PCA. Each individual step involved in our procedure is therefore carried out on a unique dataset, ensuring that findings can be transfered to applications with a typical continuous stream of input data.
As an error measure (loss function) we utilize
word error rate (WER), i.e. the ratio between errors
and total number of evaluated samples.
The genetic algorithm efficiently converges to optimal
dynamics settings with small evaluations number, see Fig. \ref{fig:pca_ga}(a),
orange dotted curve. The best obtained parameters are 
$\tau = 7 \cdot 10^{-2}, \beta=-1.69,
\Phi_0=-1.33, \rho=1.5$
with $\text{WER} = 3.8\%$. 

We keep the reservoir size at a moderate value of $N = 600$ nodes in order to
run the experiments quickly, though our hardware could support substantially
larger systems.  Our current implementation occupies about 23\% of FPGA area.
Time-multiplexing allows increasing the reservoir dimensions by only adding
more elements to the FIFO delay line, however the most of area is occupied by
circuits implementing matrix multiplication (masking and readout).  Therefore,
we estimate that using the same FPGA model and without any additional
optimization the total number of nodes can be increased up to 2500-2600 nodes.
This number, however, can be further enlarged in several orthogonal ways: (1)
by using a more powerful FPGA, which are readily available, (2) by optimizing
the hardware description code and reformulating the design directly in a lower
level language such as VHDL, (3) by even better adapting the RC concept to
existing digital hardware.


In the next step, we study the impact of dimensionality reduction
on the classification accuracy.
With the help of PCA, we decrease the number of 
input dimensions by removing the principal components
corresponding to the smallest eigenvalue magnitudes,
i.e. containing the redundant information.
General practice in PCA
is to reduce the number of principal components
so that at least 90\% of variance is preserved.
Therefore, we anticipate that 
the minimal number of principal components
that can be used with these data is four (90.9\% variance).
This hypothesis is confirmed in Fig. \ref{fig:pca_ga}(b),
where the error sharply increases
when the number of principal components goes below four.
Principal component analysis shows that
the best result
in Fig. \ref{fig:pca_ga}(b)
is obtained for 13 principal components
(97.6\% of variance). Thus, we may conclude
that the remaining $64 - 13 = 51$ principal
components carry 2.4\% of non-essential
information such as noise.
By removing those principal components we
are able to effectively filter the residual
information, which improves the recognition
accuracy (Fig. \ref{fig:pca_ga}(b)).
In the rest of our experiments
we reduce the number of input dimensions to
seven principal components,
preserving thus 94.9\% of variance.
That corresponds to 
$64 / 7 \simeq 9$
times compression rate.
In our case, where data transfer is serial,
this compression rate substantially reduces
the transmission time to FPGA processing unit.
Furthermore, read-only memory capacity, containing
coefficients for the input masking,
is reduced 9 times comparing to conventional
masking without compression. Crucially, according to Fig. \ref{fig:pca_ga}(b), recognition performance is hardly affected by this stronger compression.

The selected dimensionality reduction
consistently improves the overall accuracy
(Fig. \ref{fig:pca_ga}(a),
thick blue curve).
The parameters obtained by the GA
$\tau=7.8125 \cdot 10^{-3}$,
$\tau_D=6$, $\beta=-1.09375$, 
$\Phi_0=-3.3125$, $\rho=1.5$
result in an optimal performance of $\text{WER} = 2.1\%$.
Moreover, PCA preprocessing 
also helps the GA to converge faster.
This can be explained by the fact that
the space of sounds (and therefore, cochleagrams)
is sparse with respect to the words
pronounced by humans,
hence the majority of the
sound space is populated by
information only weakly correlated to the information content.
Therefore, a significant part of
information contained in cochleagrams
is redundant.

\begin{figure}[ht]
\centering
\includegraphics[width=\columnwidth]{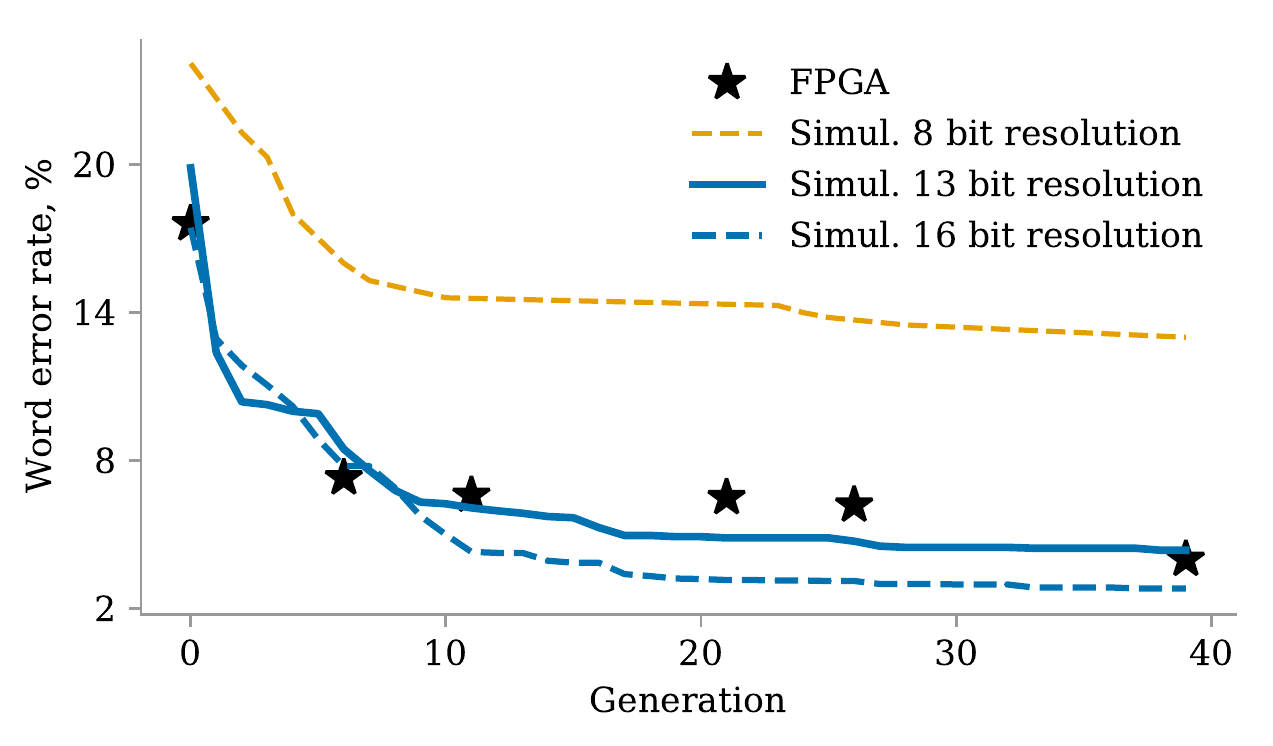}
\caption{\label{fig:rc-fpga}
\textbf{Genetic algorithm convergence
in hardware}. Simulation with noise averaged over 6 runs (lines) and FPGA realizations (stars).
The solid line corresponds to simulation with amount of noise equal to the one in FPGA.
}
\end{figure}

To better illustrate the GA search (case of PCA preprocessing), 
we perform grid search (GS) along the
three most significant parameter
dimensions, i.e. $\Phi_0$, $\beta$, and $\tau$,
crucially forced to use much coarser resolution: less than 25 points per dimension already result in a total of 11,340  points in parameter space.
The exhaustive GS in all four parameter dimensions 
with resolution comparable 
to the one we used in GA would take
5,106 times longer than GA. If, as here for our case, a single GA run in	 our implementation
takes around an hour, GS would take more than 
seven months.
Adding an additional (fifth) parameter dimension
scanned along e.g. 100 points,
would immediately increase 
the GS time to 59 years.
Both GA and GS can be parallelized, but
in case of GS, parallelization
cannot overcome the exponential growth off necessary resources.
That clearly highlights the advantage of GA over exhaustive GS.

Figure \ref{fig:pca_ga}(c) reveals error isosurfaces
in the three-dimensional parameter space
obtained as a result of GS.
The isosurfaces 
present nested objects
corresponding to WERs = 10\%, 5\%, and 3\%.
Error rates obtained from the GA search
are visualized with circles. 
It can be seen that the topmost circle
is a result of the random search 
(zero generation), corresponding $\rm{WER} \simeq 16\%$.
Then, as GA is efficiently converging,
the circles are quickly
approaching an acceptable local minimum.
Although Fig. \ref{fig:pca_ga}(c)
illustrates the GA search in only three dimensions,
GA is simultaneously optimizing parameters
in all four dimensions.

Finally, we apply the GA technique to an FPGA-based
RC. Before we actually implement RC on FPGA,
we accurately estimate
an optimal parameter set offline on a computer.
In order to take into account the quantization noise in
FPGA with limited bit resolution,
we simulate the limited bit resolution
in Eq. \eqref{eq:rc-lowpass}
corresponding to the white noise of level $2^{-13} \simeq 1.2 \cdot 10^{-4}$.
The noise is applied to the dynamical variable $x(t)$,
the delay term $x(t-\tau_D)$,
and the result of nonlinear transformation $f$.
Additionally, to better model the behavior of our hardware,
in the beginning of numerical experiment
we add the noise of the same magnitude to masking coefficients
and also we repeat the procedure with readout coefficients
right after training.

Figure \ref{fig:rc-fpga} shows the corresponding
to FPGA 13 bit quantization noise results
simulated on a computer (solid line).
To highlight the impact
of quantization noise,
we also provide simulations for
8 bit (dashed orange curve) and 16 bit (dashed blue
curve) resolutions.
Due to the noise in the experiment, the accuracy
of classification degrades overall. We see that
for the lowest, 8 bit resolution, the accuracy is significantly deteriorated.
We then select parameters from a GA optimization
under simulated quantization noise of 13 bits.
These parameters are then used for the RC implementation in the FPGA. Computational results obtained fully autonomously by the FPGA correspond to the black stars.
They excellently match the average convergence
obtained from the offline model, thereby validating our approach.


\section{Conclusion}

We have proposed a technique towards practical application
of reservoir computing (RC). The technique
consists of two components: 
data-driven input mask optimization
and efficient dynamical parameter optimization
in terms of RC evaluations number.
We have illustrated those methods and
their strong positive impact
on the speech recognition.
The advanced input masking reduced the input data
to be transfered to the device by 9 times
and lowered the average
classification error by 1.7\%, a 1.8 fold improvement.
The improved parameter optimization reduced
the number of iterative optimization steps by
5,106 times when compared to exhaustive grid search.

We took advantage of the fact that 
the exact RC model was known in advance
and were able to 
run genetic algorithm (GA)
optimizations offline on a PC.
We took into account the hardware's quantization
noise and have illustrate the significance of its impact
in possible real-world scenario.
Finally, we have built an FPGA RC
confirming our evolutionary technique.
This illustrates how our method
can be applied to
various physically existing RC
systems where noise is inevitably
present.

Another significant benefit of GA
is that the method could be applied
even when the exact model 
of the optimized system
was unknown.
This would enable RC optimization
online,
i.e. directly on the actual hardware,
as it was done e.g. in \cite{Thompson1996}.
In this work we have shown that evolutionary-inspired
optimization can significantly reduce
the time to adapt RC dynamics to an unforeseen task.
We leave the implementation of
online GA to the future investigations
as the next logical step towards self-adapting hardware \cite{Aporntewan2001AAlgorithm}.

\section*{Acknowledgments}

This work was supported by 
the Labex ACTION program (Contract No. ANR-11-LABX-0001-01)
and by the BiPhoProc ANR project (ANR-14-OHRI-0002-02).

\end{document}